\documentclass[preprint,12pt]{elsarticle}

\usepackage{amssymb}
\usepackage{amsthm}
\usepackage{amsmath}
\usepackage[colorlinks=true, urlcolor=blue, linkcolor=blue, citecolor=blue]{hyperref}
\usepackage{xcolor}
\usepackage{float}
\usepackage{lineno}
\usepackage{tikz}
\usepackage{xfrac}
\usepackage{setspace}
\usetikzlibrary{positioning, arrows.meta, shapes.geometric, calc}

\journal{Nuclear Instruments A}

\begin{document}

\begin{frontmatter}



\title{AC-LGADs Fermilab Front-End Electronics Characterization}


\author[ULS]{René Ríos}
\author[USM,CCTVAL]{Esteban Felipe Molina Cardenas}
\author[FERMILAB]{Cristian Peña}
\author[ULS,SAPHIR]{Orlando Soto}
\author[SAPHIR,USM,CCTVAL]{William Brooks}
\author[FERMILAB]{Artur Apresyan}
\author[FERMILAB]{Sergey Los}
\author[USM,CCTVAL]{Claudio San Martín}

\affiliation[ULS]{
    organization={Departamento de Física, Universidad de La Serena},
    addressline={Av. Juan Cisternas 1200}, 
    city={La Serena},
    state={Coquimbo},
    country={Chile}}

\affiliation[SAPHIR]{
    organization={Millennium Institute for Subatomic Physics at the High-Energy Frontier (SAPHIR) of ANID},
    addressline={Fernández Concha 700}, 
    city={Santiago},
    country={Chile}}
\affiliation[USM]{
    organization={Departamento de Física y Astronomía, Universidad Técnica Federico Santa María},
    city={Valparaíso},
    country={Chile}}
\affiliation[CCTVAL]{
    organization={Centro Científico Tecnológico de Valparaíso-CCTVal, Universidad Técnica Federico Santa María},
    city={Valparaíso},
    postcode={Casilla 110-V},
    country={Chile}}
\affiliation[FERMILAB]{
    organization={Fermi National Accelerator Laboratory},
    addressline={PO Box 500}, 
    city={Batavia},
    postcode={60510-5011},
    state={IL},
    country={USA}}

\begin{abstract}
We characterized the front-end electronics used to process high-frequency signals from low-gain avalanche diodes (LGADs) at the Fermilab Test Beam Facility. LGADs are silicon detectors employed for charged particle tracking, offering exceptional spatial and temporal resolution. The purpose of this characterization was to understand how the time resolution is influenced by the front-end electronics. To achieve this, we developed a setup capable of generating input signals with varying amplitudes. The output results demonstrated that signal processing by the front-end electronics plays a crucial role in enhancing time resolution. We showed that the time resolution achieved by the FEE board is better than $2\: ps$ at the $1\sigma$ level.
\end{abstract}


\end{frontmatter}


\section{Introduction}
The luminosity upgrade of the LHC brings new challenges: up to hundreds of simultaneous collisions per bunch crossing are expected with this upgrade \cite{Sicking,penaPrecisionTimingCMS2019,Wulz_2015}; therefore, it is imperative to distinguish these events from each other to ensure a good quality particle identification (PID) and vertex reconstruction.\\
One of the alternatives to take on these challenges is the AC-coupled low-gain avalanche diode (AC-LGAD), a silicon-based device that provide an excellent reconstruction of spatial and time information \cite{Giacomini:2019kqz,Mandurrino2019,Mandurrino2020}. Recent studies \cite{hellerCharacterizationBNLHPK2022} estimate a spacial and time resolution in the order of 6-10 $\mathrm{\mu}$m and 30 ps respectively. \\

\begin{figure}[h!]
    \centering
    \includegraphics[width=8cm]{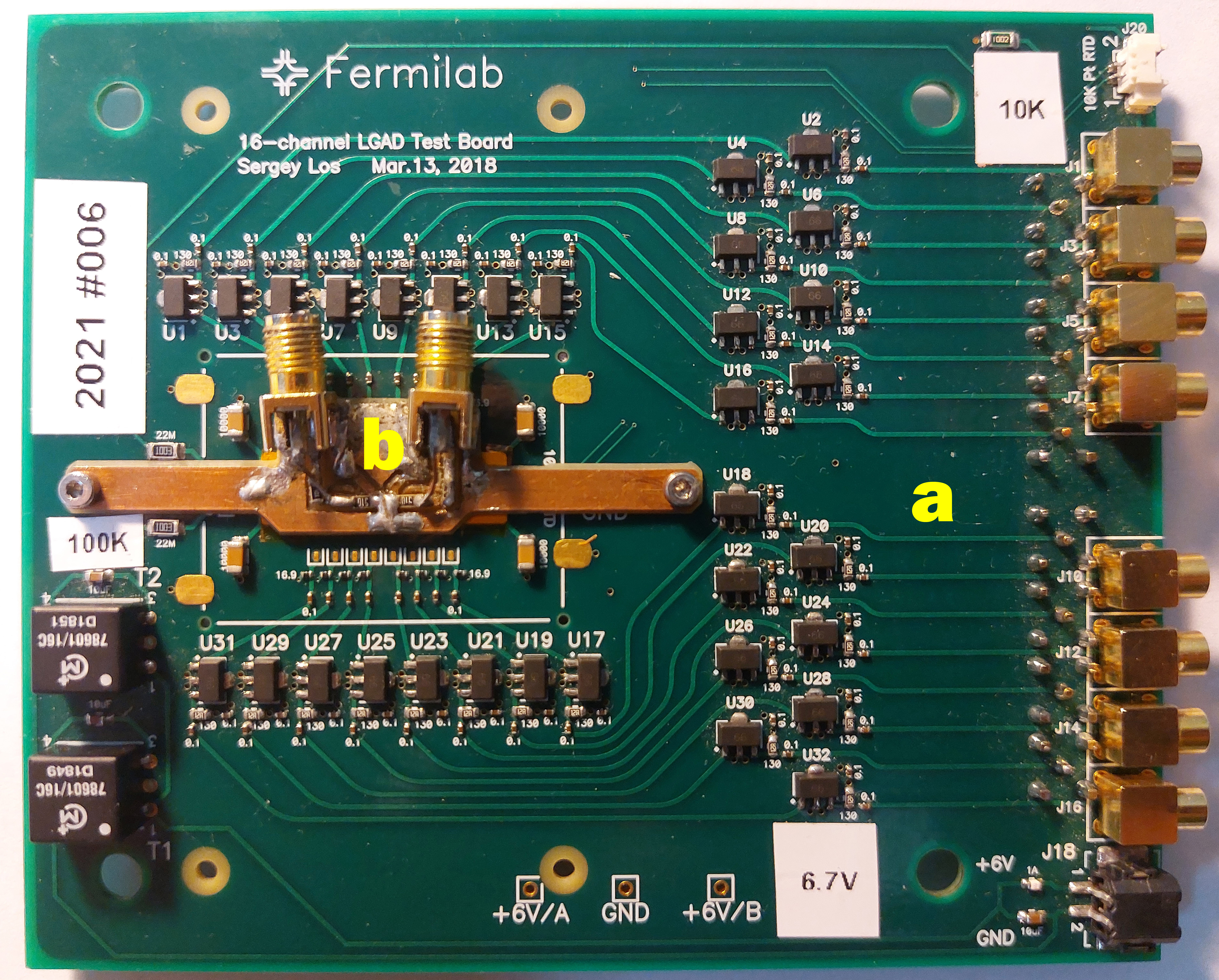}
    \caption{FEE board (a) with the designed charge injector (b) in place of an AC-LGAD.}
    \label{fig:board}
\end{figure}

In order to extract the performance limits of the AC-LGAD, a specially designed front end electronics (FEE) must be used. A 16-channel LGAD Test board (Fig.\ref{fig:board}) was used as FEE to amplify and read the signals acquired by the AC-LGADs. This board was utilized successfully in the 2021 \cite{hellerCharacterizationBNLHPK2022} and 2022 \cite{Madrid_2023} AC-LGADs test beam campaigns at the Fermilab Test Beam Facility.
The main objective of this research is the characterization of the Fermilab FEE Board, which acts as the Device Under Test (DUT), starting from detailing its electronic schematic, the experimental setup and equipment utilized, the features of the signals used to characterize the DUT, and the results in terms of frequency response, gain, noise RMS, signal-to-noise ratio (SNR), and time resolution (jitter).

\section{AC-LGADs Fermilab Front-End Electronics}

\begin{figure}[h!]
    \centering
    \includegraphics[width=12cm]{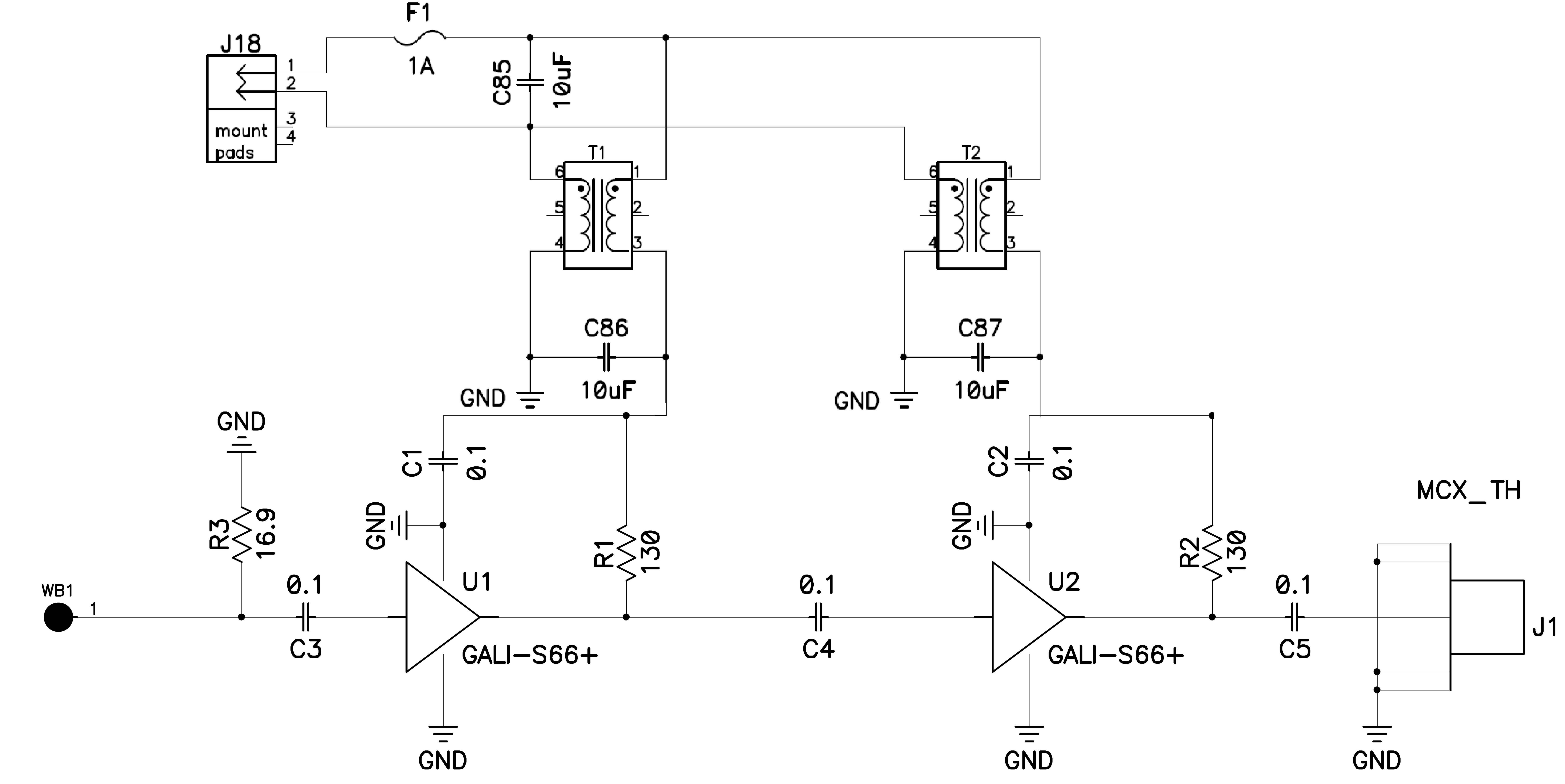}
    \caption{Single channel equivalent schematic.}
    \label{fig:circuit}
\end{figure}

The DUT has $16$ dedicated input channels for each strip or pad of the corresponding AC-LGAD device. Each channel consists of a two-stage amplifier based on the \textit{Minicircuits Gali S66+} surface mount monolithic amplifier, which provides amplification in the $DC$ to $3\: \rm GHz$ frequency range. The amplifier is internally matched to $50\: \Omega$, which requires that the traces of the board are designed with a $50\: \Omega$ impedance. The schematic of a single channel is shown in Fig. \ref{fig:circuit}. In this particular configuration, the amplifiers used a $25\:\Omega$ input impedance and a bandwidth of $1\: GHz$. The amplifier chain of each readout channel has a uniform gain with an approximate $10\%$ variation from channel to channel \cite{hellerCombinedAnalysisHPK2021}.
At $2\: GHz$, the expected gain for each amplifier stage is about $18.2\: dB$. The manufacturer specifies \cite{SMTLowNoise} the typical output power at $1\: dB$ compression is $3.3\: dBm$, at $2\: GHz$. This output, at $50\: \Omega$, is equivalent to $\sim 327\: mV$. Due to having two stages, the gain at $2\: GHz$ is a factor of $\sim66$, hence a voltage range amplitude in $0 \: -\: 4.95\: mV$ ensures a constant gain at that frequency. The total trans-impedance on the DUT, after two stages of amplification, is roughly $4.3\: k\Omega$. The gain for LGAD signals is such that each femtocoulomb of input charge results in an additional $5\:mV$ in signal amplitude (that is, $100\:mV$ output for a $20 fC$ input) \cite{hellerCharacterizationBNLHPK2022}.\\

\section{Experimental Setup Overview}
\begin{figure}[H]
    \centering
    \includegraphics[width=10cm]{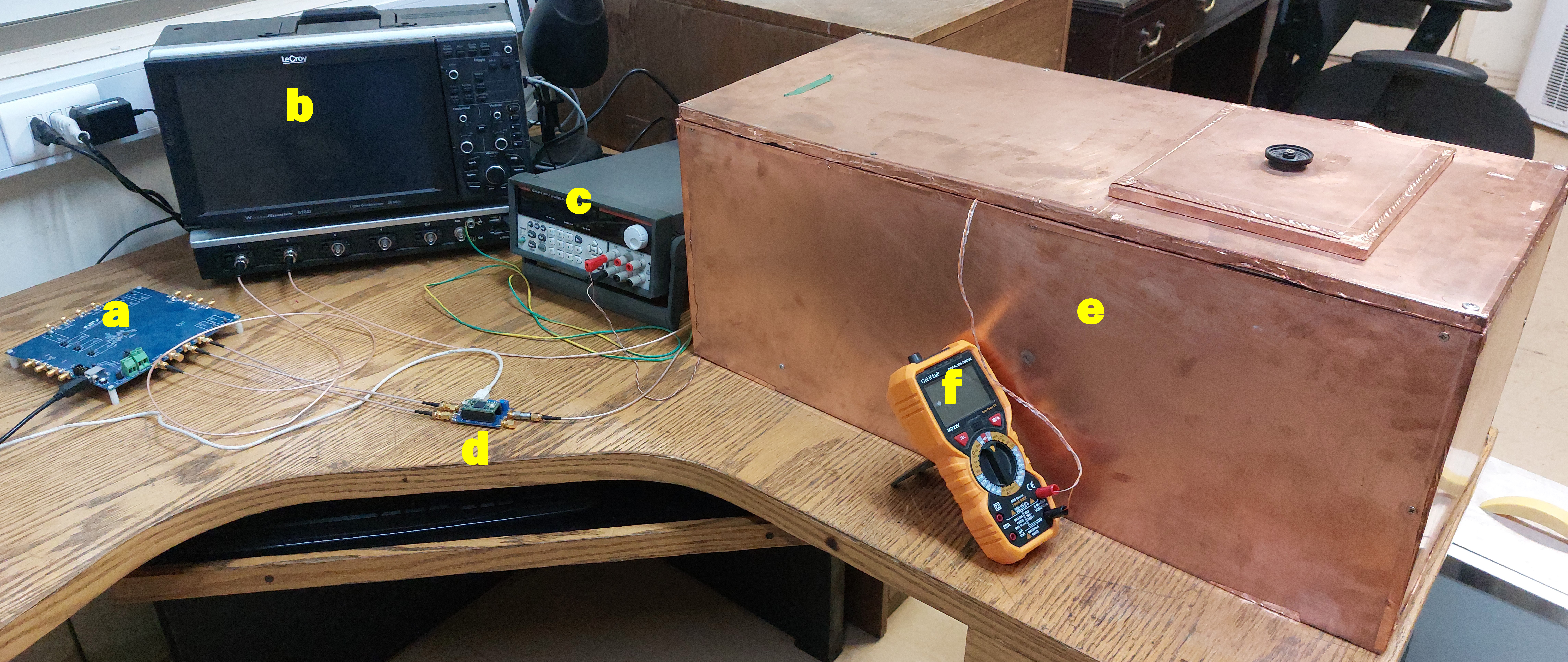}
    \caption{From left to right: clock generator (a), oscilloscope (b), power supply (c), pulse generator (d), EMI shielded enclosure containing the DUT (e), and multimeter for RTD measurement (f).}
    \label{fig:setup-box}
\end{figure}

For the characterization process, we set basic conditions regarding the powering of the board and its temperature. We set $6.8\: V$ in a \textit{Keithley 2230-30-1 DC Power Supply} to power the DUT. At that voltage, the current required by the DUT was $0.537\: A$. The board was kept inside the box (Fig. \ref{fig:setup-box}) at a temperature of $\sim 21.7\: ^\circ C$ throughout the tests using a \textit{Lauda Alpha RA 8} recirculating chiller. We measured the temperature with a $10K$ RTD sensor connected directly to the DUT.

\begin{figure}[H]
    \centering
    \includegraphics[width=12cm]{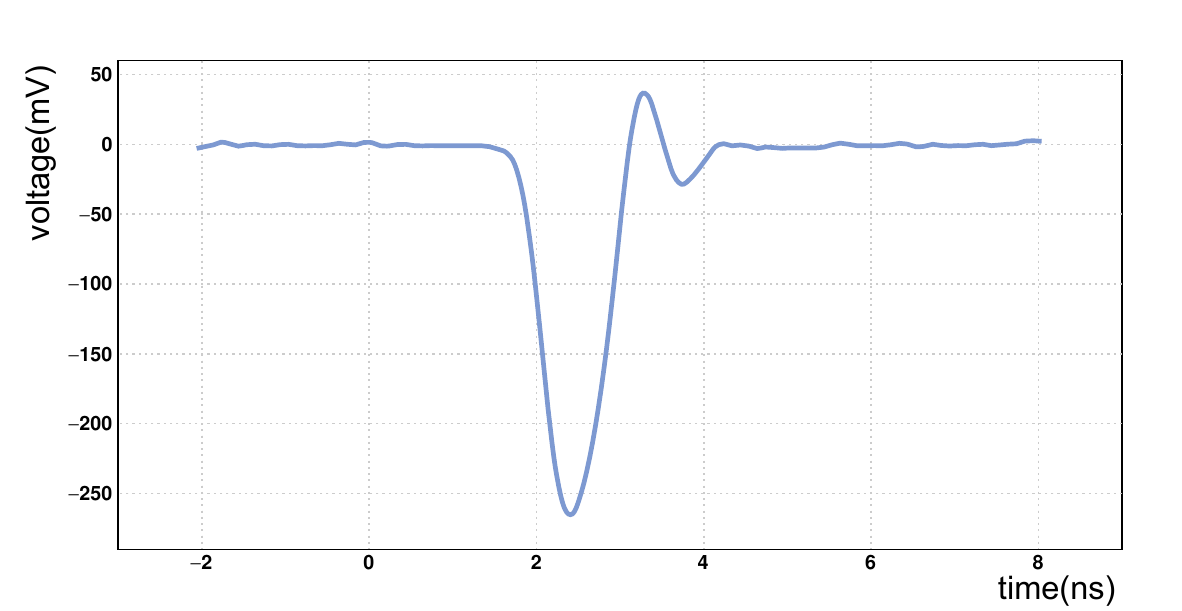}
    \caption{Output from the pulse generator without any attenuation applied.}
    \label{fig:pg-signal}
\end{figure}

We used a \textit{SiLabs Si5332-6EX-EVB REV 2.0} LVDS low-jitter clock generator (less than 175 fs RMS phase jitter) configured at $84.3\:MHz$ to produce squared signals of fixed width and amplitude. One of the outputs of the clock generator is connected directly to the first channel of the oscilloscope as a trigger for the experiment. Two other outputs of the clock generator fed a pulse generator, designed by the CALTECH INQNET \cite{FQNETHome} team, based on the \textit{NB6L295 Dual Channel Programmable Delay Chip}. This device takes two input signals and produces two delayed signals which are driven into a \textit{MC100LVEP05} Low Voltage Positive Emitter-Couple Logic (LVPECL) \textit{AND} gate. The differences in the delays can produce nanosecond-width pulses, that would emulate those produced by a Minimum Ionizing Particle (MIP) when hitting an AC-LGAD. Additionally, we used different attenuation levels at the output of the pulse generator to achieve a similar voltage amplitude to those produced by the signals of MIPs and to study the behavior of the SNR.
The range of amplitudes for the input test pulses was chosen to be representative of the charge deposited by a Minimum Ionizing Particle (MIP) in a 50~$\mu$m thick AC-LGAD. For such detectors, a collected charge of approximately 20~fC is a typical operating point to achieve a time resolution of 30~ps \cite{hellerCombinedAnalysisHPK2021}. The FEE board under study provides a gain such that a 20~fC input charge results in a 100~mV output signal \cite{hellerCharacterizationBNLHPK2022}. Considering the board's voltage gain of approximately 66 (a factor derived from its two-stage amplification), this corresponds to an input pulse amplitude of about 1.5~mV. As detailed in Table~\ref{tab:gain_dB}, the input pulse amplitudes used in this work range from 0.942~mV to 5.189~mV, which effectively covers the signal range relevant for MIP detection and allows for the characterization of the SNR performance around this critical operating point.
The signal produced by the pulse generator (see Fig. \ref{fig:pg-signal}) was set at $1\:ns$ width and $260\:mV$ amplitude. The measured 20-80\% rise time of the output pulses is on average $\sim350$~ps, a value that is primarily limited by the instrument response of the 1~GHz bandwidth oscilloscope. The non-ideal pulse shape, including the observed overshoot and subsequent ringing, is a characteristic artifact of the pulse generator's design, which relies on a high-speed LVPECL logic gate (\textit{MC100LVEP05}) not originally intended for generating clean analog test signals. To provide a single-ended input to the charge injector, one of the two differential outputs of the pulse generator is used, which is then AC-coupled to remove its DC component.

We manufactured a charge injector (Fig. \ref{fig:board}) to guide the output signals of the pulse generator into an input channel of the DUT. The injector consisted of an SMA input connector with a $50\: \Omega$ resistor coupled to ground to match the input impedance of the cables; additionally, a $100\: nF$ capacitor was connected in series with the input connector to eliminate LVPECL common mode voltage, and at the output of the charge injector, we used a spring-loaded contact connector. To isolate the DUT from external electromagnetic noise, it was placed inside a custom-built EMI shielded enclosure (Fig.~\ref{fig:setup-box}). The data acquisition was performed using a \textit{LeCroy WaveRunner 610Zi} oscilloscope. This instrument has a $1\:GHz$ bandwidth and a maximum sampling rate of $20\:GSa/s$. The bandwidth imposes a manufacturer-specified $10-90\%$ rise time of $375\: ps$, which sets a fundamental limit on the measurement of fast signal transitions.

We arranged these devices in different configurations to perform the characterization.
\section{Gain and Jitter characterization}
\subsection{Experimental Setup} 

\begin{figure}[htbp!]
	\centering
	\begin{tikzpicture}[
		node distance=0.6cm,
		block/.style={rectangle, draw, minimum width=1cm, minimum height=0.8cm, align=center},
		arrow/.style={-{Latex}, thick},
		font=\ttfamily\scriptsize
		]
		
		\node[block] (clock) {Clock}; 
		\node[block, right=of clock] (pulse) {Pulse\\Generator};
		\node[block, right=of pulse] (attenuator) {Attenuator};
            \node[block, right=of attenuator,
                text width=3cm,
                align=center,
                path picture={
                    \draw[dashed]
                    ([xshift=0cm]path picture bounding box.north) -- 
                    ([xshift=0cm]path picture bounding box.south);
                }
            ] (readout) {
                \begin{tabular}{p{1cm}  p{1.3cm}}
                \centering Charge\\Injector & \centering FEE\\Board\\(DUT)
                \end{tabular}
            };
		\node[block, right=of readout] (oscilloscope) {Oscilloscope};
		
		\draw[arrow] (clock) -- (pulse);
		\draw[arrow] (clock.north) to[out=25, in=155] node[pos=1,above]{\tiny \hspace*{0.2cm} Clock} (oscilloscope.north);
		\draw[arrow] (pulse) -- (attenuator);
		\draw[arrow] (attenuator) -- (readout);
		\draw[arrow] (readout) -- node[pos=1,above]{\tiny Pulse \hspace*{0.5cm}} (oscilloscope);

	\end{tikzpicture}
	\caption{Experimental setup for amplitude, noise RMS and jitter analysis.}
	\label{fig:exp-setupA}
\end{figure}
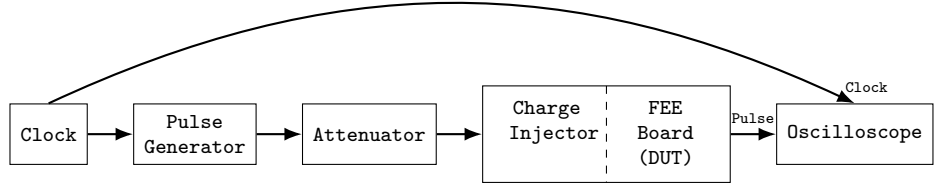

The experimental setup used for the gain and jitter characterization is depicted in Fig.~\ref{fig:exp-setupA}. A low-jitter clock generator provides the primary time reference for the system. This clock signal is sent to two destinations: directly to a reference channel on the oscilloscope, and as a trigger to the pulse generator. The pulse generator produces test signals that emulate those from a MIP. These signals are then passed through a variable attenuator, which allows for the adjustment of their amplitude to study the DUT's performance across a range of input signal levels. The attenuated pulse is fed into the FEE board (DUT) via the custom charge injector. Finally, the oscilloscope simultaneously captures the amplified output signal from the DUT and the original reference clock signal. This configuration allows for a precise measurement of the output signal's characteristics (amplitude, rise time) and the time difference between the reference and the DUT output, which is essential for the jitter analysis.

\subsection{Overview of The Pulses}
\begin{figure}[H]
	\centering
	\includegraphics[width=12cm]{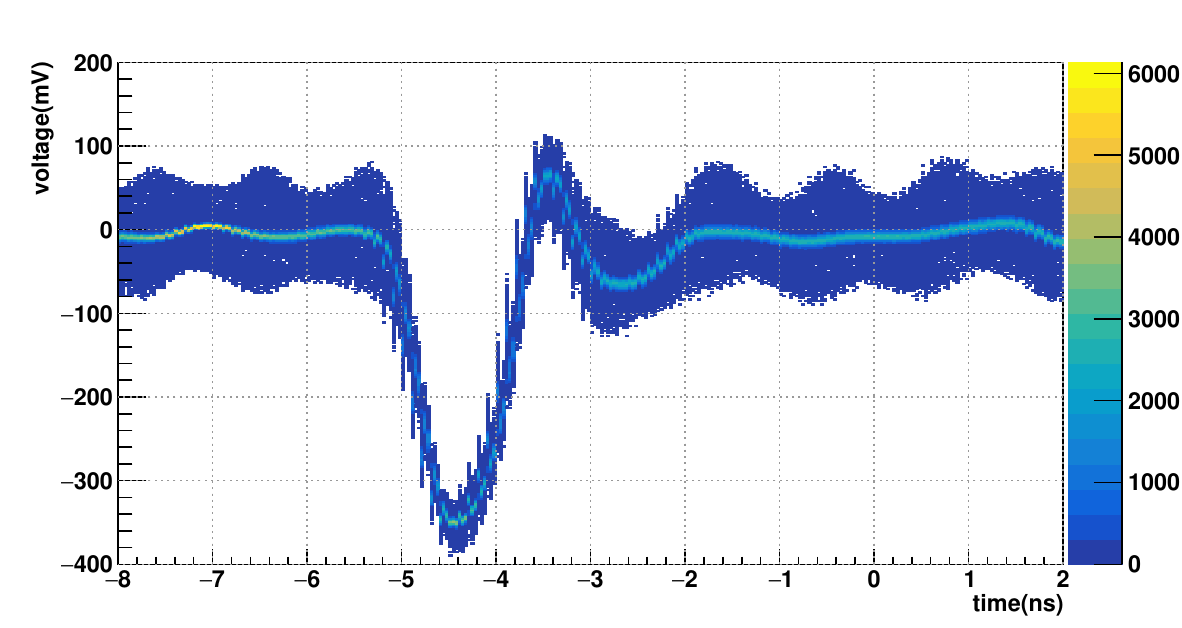}
	\caption{5000 events acquired from the output of the readout board of FNAL using the CMS-MTD Timing DAQ codes \cite{TimingDAQRepository2022}.}
	\label{fig:pulse-overview}
\end{figure}

\begin{figure}[H]
	\centering
	\includegraphics[width=12cm]{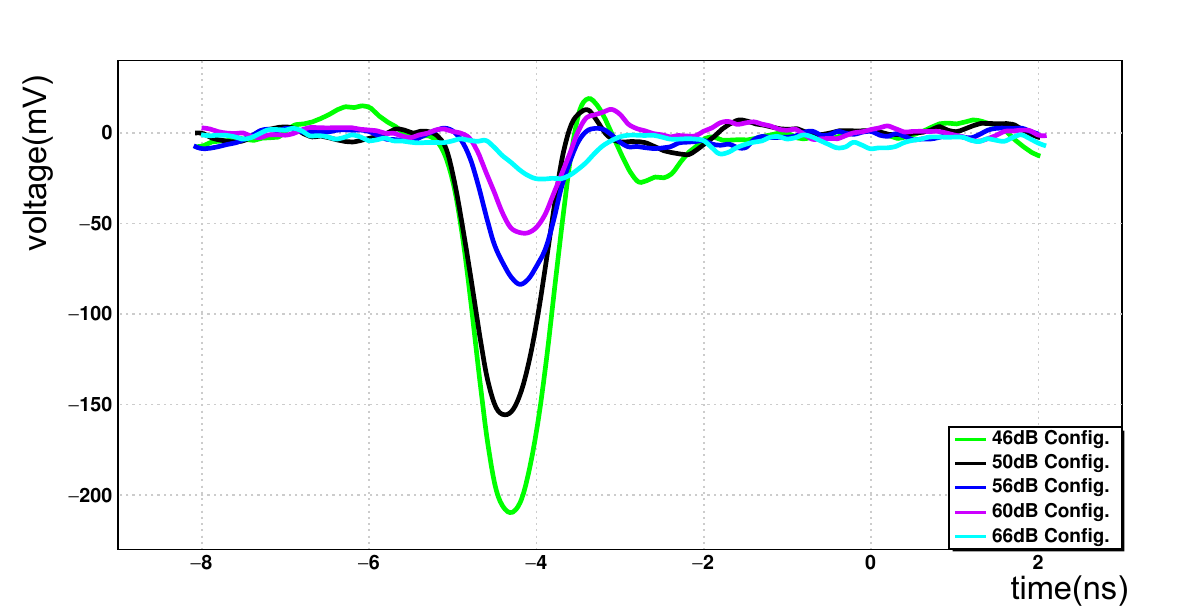}
	\caption{Outputs from the FEE for different attenuators between the pulse generator and the FEE input.}
	\label{fig:configs-comparison}
\end{figure}

We used specially-developed software to acquire and convert binary data directly from the oscilloscope to ROOT framework data files  \cite{brunRootprojectRootV62019}. This software was developed by the FNAL CMS-MTD team, which was successfully utilized in previous works \cite{hellerCharacterizationBNLHPK2022,Madrid_2023}. As a reference, Fig. \ref{fig:pulse-overview} shows $5000$ events for the $40\:dB$ configuration. 
In Fig. \ref{fig:configs-comparison} can be seen a comparative plot between one event for every configuration.

\subsection{Gain Results}

\begin{figure}[H]
    \centering
    \includegraphics[width=10cm]{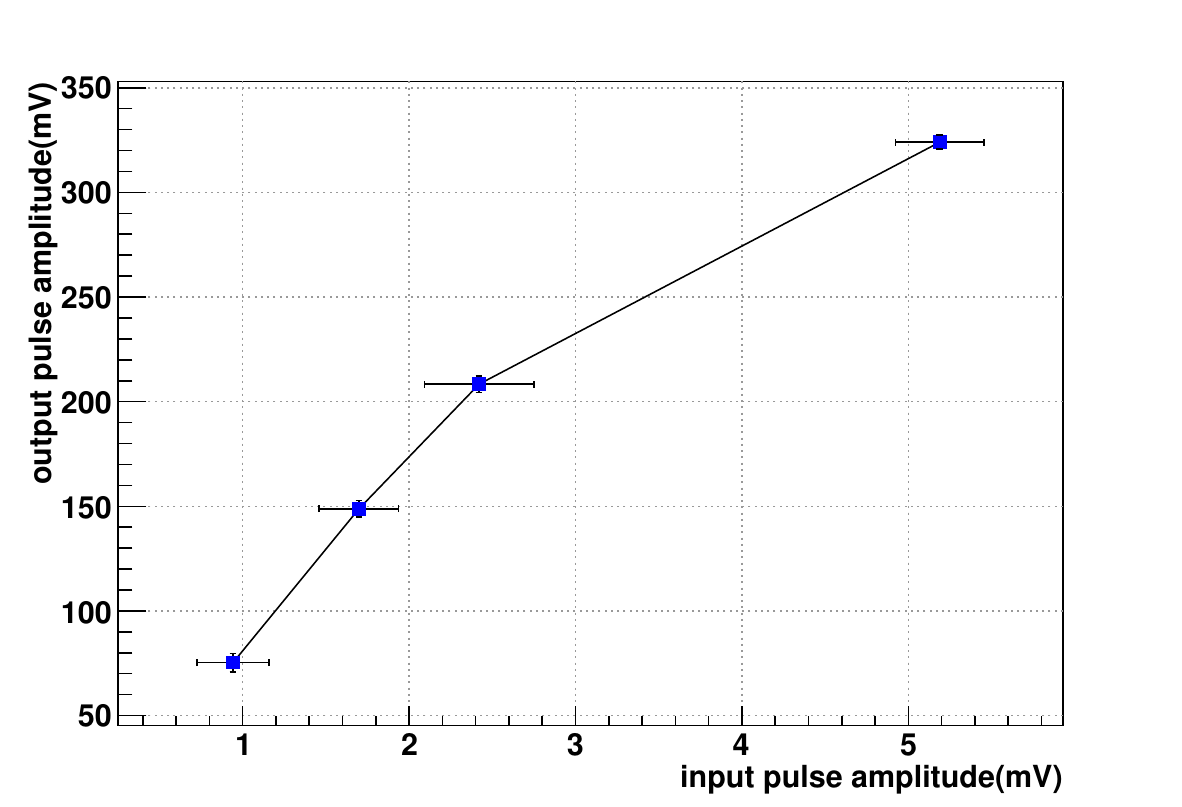}
    \caption{Output amplitudes as a function of input amplitudes, channel $8$.}
    \label{fig:in_out_pulse}
\end{figure}

Pulses of different amplitudes were applied at the input pad of channel $8$ of the board, with the clock generator set at $84.3\: MHz$. The output amplitudes were measured as seen in Fig. \ref{fig:in_out_pulse}.

As a result from the amplitude comparison, the gain was calculated for each pulse, as presented in table \ref{tab:gain_dB}. This was corroborated with the VNA results shown in the following section. 

The output signals were characterized over a wide range of input attenuations, from 40~dB to 66~dB, as shown in Fig.~\ref{fig:configs-comparison}. However, direct measurement of the corresponding \textit{input} pulse amplitudes was only feasible for attenuations up to 56~dB. For higher attenuation values (i.e., 60~dB and 66~dB), the input signal amplitude fell below the noise floor of the oscilloscope, preventing a reliable characterization. For this reason, the quantitative gain analysis presented in Table~\ref{tab:gain_dB} and Fig.~\ref{fig:in_out_pulse} is limited to the subset of measurements where both input and output signals could be accurately measured.
\begin{table}[H]
	\centering
	\small
	\begin{tabular}{|c|c|c|c|}
		\hline
		\multicolumn{1}{|l|}{\textbf{\begin{tabular}[c]{@{}l@{}}Attenuation at pulse \\ generator output dB\end{tabular}}} & \multicolumn{1}{l|}{\textbf{\begin{tabular}[c]{@{}l@{}}Input pulse\\ amplitude mV\end{tabular}}} & \textbf{\begin{tabular}[c]{@{}c@{}}Output pulse\\ amplitude mV\end{tabular}} & \multicolumn{1}{l|}{\textbf{Gain dB}} \\ \hline
		\textbf{56} & $0.942 \pm 0.217$ & $75.267 \pm 4.459$ & $38.05 \pm 2.07$ \\ \hline
		\textbf{50} & $1.698 \pm 0.238$ & $148.785 \pm 4.008$ & $38.85 \pm 1.24$ \\ \hline
		\textbf{46} & $2.421 \pm 0.328$ & $208.359 \pm 3.888$ & $38.70 \pm 1.19$ \\ \hline
		\textbf{40} & $5.189 \pm 0.267$ & $324.017 \pm 3.392$ & $35.91 \pm 0.47$ \\ \hline
	\end{tabular}
	\caption{Output amplitudes from the FEE and the corresponding input pulse amplitude at the channel input. The attenuators are installed after the pulse generator to get the shown input amplitudes. The drop in the gain for the biggest input pulse can be explained by the gain compression of the \emph{Minicircuits Gali S66+} amplifier at that amplitude.}
	\label{tab:gain_dB}
\end{table}

\subsection{Jitter measurement}

To evaluate the time resolution, the signals were processed using the CMS-MTD Timing DAQ framework \cite{TimingDAQRepository2022}. After identifying the rising edge on the selected oscilloscope input channel, the time difference between input signals was computed at a constant fraction of their respective amplitudes. The variance of the resulting time difference distribution, as defined in Equation~\ref{eq:delta_T}, corresponds to the measured jitter. Using the experimental setup shown in Fig.~\ref{fig:exp-setupA}, the analysis focused on the time difference between a reference signal (clock) and the device under test (FEE board).

\begin{equation} \label{eq:delta_T}
	\Delta t = t(50\%\text{ Rising Edge @ Reference}) - t(50\%\text{ Rising Edge @ DUT}),
\end{equation}

\subsubsection{Time resolution}
If a noisy analog pulse is applied to a leading edge trigger, the timing uncertainty can be obtained by projecting the variance $\sigma_n$ of the momentary signal amplitude on its rate of change at the trigger threshold $V_T$. This yields the variance in time $\sigma_t$ -time resolution-, called jitter \cite{Spieler:1982jf}. 
\begin{equation}
     \sigma_t=\frac{\sigma_n}{\frac{dV}{dt}\big\rvert_{V_T}}+ \delta t
\end{equation}

Assuming the rising edge is roughly described by a straight line, the previous expressions can be associated with the following concepts:
\begin{itemize}
    \item $\sigma_n$ to the  \text{Noise RMS}.
    \item $\sigma_t$ to the  \text{Time Resolution}.
    \item $\frac{dV}{dt}$ to $\frac{\text{Amplitude}}{\text{Rise Time}}$.
    \item $\delta t$ to the the residual jitter of the associated electronics.
\end{itemize}

We can project the noise on its rate of change at the trigger threshold to obtain an expression for the jitter \cite{Spieler:1982jf}. Thus, re-arranging the previous approximations we obtain:
\begin{equation}\label{eq:time_resolution_analytic}
    \sigma_t \approx \frac{\text{Rise Time}}{\text{SNR}}+\delta t
\end{equation}

Hence, an increase in the SNR should provide a better time resolution, since the rise time depends on fixed characteristics of the input signal, readout electronics and ADC bandwidth. Moreover, we notice that the last term cannot be reduced as we increase the input signal, so the asymptotic behavior of the jitter $\delta t$ will be the \textit{time resolution of the system}. Thus, as part of the characterization of the FEE, we will measure the contribution of the FEE readout board in the time resolution to understand its impact on the AC-LGADs measurements.\\
We set the same voltage and time scales in the oscilloscope for most of the configurations since a variation in the voltage scale can change the level of ADC noise.

\subsubsection{Pulse generator and associated electronics jitter }

\begin{figure}[htbp!]
	\centering
	\begin{tikzpicture}[
		node distance=0.6cm,
		block/.style={rectangle, draw, minimum width=1cm, minimum height=0.8cm, align=center},
		arrow/.style={-{Latex}, thick},
		font=\ttfamily\scriptsize
		]
		
		\node[block] (clock) {Clock}; 
		\node[block, right=of clock] (pulse) {Pulse\\Generator};
		\node[block, right=of pulse] (readout) {Charge\\Injector};
		\node[block, right=of readout] (oscilloscope) {Oscilloscope};
		
		\draw[arrow] (clock) -- (pulse);
		\draw[arrow] (clock.north) to[out=25, in=155] node[pos=1,above]{\tiny \hspace*{0.2cm} Clock} (oscilloscope.north);
		\draw[arrow] (pulse) -- (readout);
		\draw[arrow] (readout) -- node[pos=1,above]{\tiny Pulse \hspace*{0.5cm}} (oscilloscope);

	\end{tikzpicture}
	\caption{Experimental setup for jitter analysis of the pulse generator and associated electronics. The attenuators were not considered in the analysis since their effect in the time resolution is negligible and they were only used to change the input pulse amplitudes.}
	\label{fig:exp-setupA2}
\end{figure}
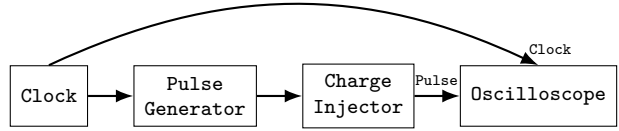

As part of the characterization, we measured the jitter of the system comprising the clock, pulse generator, charge injector, and oscilloscope, as shown in Fig. \ref{fig:exp-setupA2}. To facilitate this measurement, a custom adapter was developed to enable a direct connection between the oscilloscope and the charge injector.

To measure the baseline jitter of the pulse generator and oscilloscope ($\sigma_{pulse\:generator}$), the output of the generator was fed into a passive 50~$\Omega$ power splitter. The two identical resulting signals were connected to two separate channels of the oscilloscope. The jitter was then calculated from the standard deviation of the distribution of the time difference between the two channels, measured at a constant fraction of their amplitude. This differential measurement method effectively cancels the intrinsic jitter of the pulse generator, isolating the combined jitter contributions from the oscilloscope and cabling, which constitutes the baseline resolution of the measurement system.

The measured time resolution is given by:

\begin{equation}\label{eq:time_resolution_PG}
\sigma_{pulse\:generator}= 4.13 \pm 0.12\: (ps)    
\end{equation}

\subsubsection{Output amplitudes, rise time, jitter, noise and SNR}

Using the setup shown in Figure~\ref{fig:exp-setupA}, we characterized several parameters of the output signals. The results are summarized in Table~\ref{tab:noise-RMS}. An increase in the signal-to-noise ratio (SNR) is observed to significantly enhance the time resolution, highlighting the importance of front-end electronics (FEE) in the performance of such sensors. Notably, the noise levels remain approximately constant across different output amplitudes, underscoring that the improvement in resolution is primarily driven by the increase in signal amplitude rather than a reduction in noise. It should be noted that the measured rise times are comparable to the oscilloscope's manufacturer-specified $10-90\%$ rise time of $375\: ps$. Therefore, the values presented in the table should be considered an upper limit, as they are likely dominated by the instrumental response of the oscilloscope.\\

\begin{table}[h!]
\centering
\small
\begin{tabular}{|c|c|c|c|c|c|}
\hline
\textbf{Att.} & \textbf{Rise} & \textbf{Noise} & \textbf{Amp.} & \textbf{SNR} & \textbf{Jitter} \\
\textbf{(dB)} & \textbf{$\text{time }_{20-80\%}$ (ps)}  & \textbf{RMS (mV)} & \textbf{(mV)} & & \textbf{(ps)} \\
\hline
36 & $300 \pm 6$   & $1.84 \pm 0.40$  & $795 \pm 1$     & $487 \pm 105$ & $4.63 \pm 0.03$ \\
40 & $344 \pm 14$  & $3.04 \pm 1.04$  & $331 \pm 3$     & $129 \pm 33$  & $6.14 \pm 0.11$ \\
46 & $353 \pm 16$  & $2.57 \pm 0.86$  & $206 \pm 4$     & $76 \pm 21$ & $8.21 \pm 0.16$ \\
50 & $382 \pm 28$  & $2.29 \pm 0.85$  & $152 \pm 3$     & $82 \pm 22$ & $11.36 \pm 0.17$ \\
56 & $358 \pm 40$  & $2.89 \pm 1.01$  & $73 \pm 4$      & $47 \pm 14$ & $22.66 \pm 0.30$ \\
60 & $435 \pm 59$  & $2.20 \pm 0.87$  & $53 \pm 4$      & $31 \pm 9$  & $30.38 \pm 0.46$ \\
66 & $357 \pm 94$  & $2.38 \pm 0.94$  & $25 \pm 4$      & $13 \pm 4$  & $78.55 \pm 1.39$ \\
\hline
\end{tabular}
\caption{Summary of the extracted parameters. Reported uncertainties are statistical only and do not account for systematic effects. }
\label{tab:noise-RMS}
\end{table}

\subsubsection{FEE Board Time Resolution}

\begin{figure}[H]
    \centering
    \includegraphics[width=10cm]{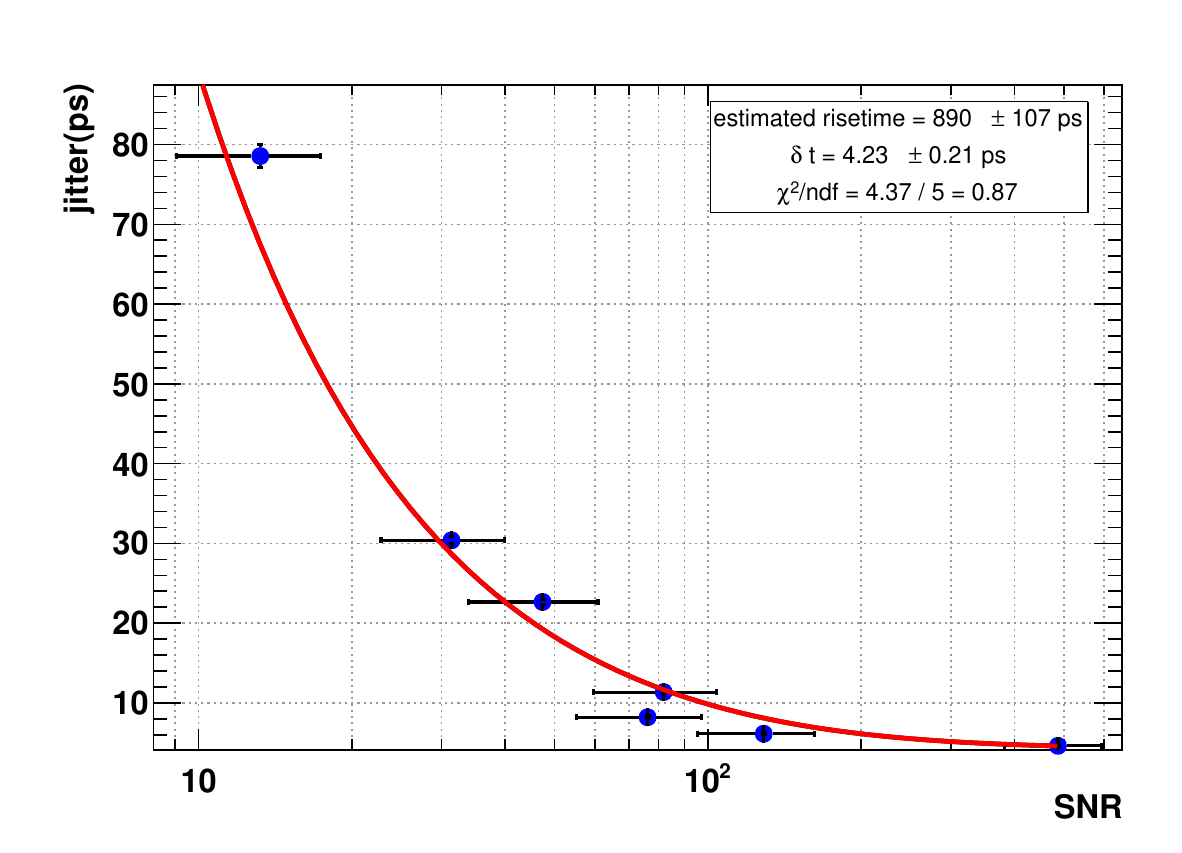}
\caption{Jitter as a function of signal-to-noise ratio (SNR). The solid line represents a weighted least-squares fit to the data using a model that quadratically adds an SNR-dependent term and a constant asymptotic jitter floor ($\delta t$). The resulting fit parameters, including the extracted asymptotic jitter, are displayed.}
    \label{fig:resolution-SNR}
\end{figure}

As demonstrated in the model proposed by~\cite{Spieler:1982jf}, the time resolution exhibits an asymptotic behavior as a function of the signal-to-noise ratio (SNR), a trend that is clearly observed in Fig.~\ref{fig:resolution-SNR}.

The data were fitted using a model based on the quadratic sum of an SNR-dependent term and a constant noise floor, $\delta t$, given by the function $\sqrt{\left(\sfrac{risetime}{SNR}\right)^2 + \delta t^2}$. This model provides an excellent description of the data, as indicated by the resulting $\sfrac{\chi^2}{ndf}=0.87$.

The fit yields an asymptotic jitter of $\delta t = 4.23 \pm 0.21\ ps$. The first parameter represents an effective rise time, and is found to be $890 \pm 107\ ps$. It is important to note that this effective parameter should not be directly identified with the physically measured $20-80\%$ rise time of the individual pulses (which is, on average, $\sim 350\ ps$ according to Table~\ref{tab:noise-RMS}). Instead, this fit parameter absorbs more complex system behaviors, including the slight variation of the rise time with SNR, into a single effective value that best describes the overall trend of the data. The primary result from this robust fit is the extraction of the physically consistent asymptotic jitter floor, $\delta t$.

Assuming that the jitter introduced by the FEE board is statistically uncorrelated with the jitter contributions from other elements in the experimental setup, the total time resolution can be expressed as:

\begin{equation}\label{eq:rb-resolution}
    \delta t^2 = \sigma^2_{FEE} + \sigma^2_{pulse\:generator}
\end{equation}
where:
\begin{itemize}
    \item $\sigma_{FEE}$ is the time resolution of the FEE board.
    \item $\delta t$ is the time resolution of the experimental setup in Fig. \ref{fig:exp-setupA}.
    \item $\sigma_{pulse\:generator}$ is the time resolution of the experimental setup in Fig.  \ref{fig:exp-setupA2}.
\end{itemize}

Using this relation and the previously quoted values, the time resolution of the readout board was estimated as:
\begin{equation} \label{eq:time_res}
\sigma_{\text{FEE}} = 0.91 \pm 1.12\: (ps)    
\end{equation}

Since the uncertainty is larger than the central value, the result is consistent with zero. We therefore set an upper limit on the contribution of the FEE board to the time resolution of $2\:ps$ ($1\sigma$ confidence level).

\section{Frequency Response}

\subsection{Experimental Setup}

We used a 6~GHz bandwidth vector network analyzer (VNA), branded \textit{NanoRFE VNA6000}, to measure the frequency response ($S_{21}$) of a single channel. To protect the VNA's ports from the high gain of the DUT, a $20\:dB$ attenuator was placed at the input of the charge injector and a $10\:dB$ attenuator at the output of the FEE board. The complete signal path for the measurement was: VNA Port~1 $\rightarrow$ 20~dB Attenuator $\rightarrow$ Charge Injector $\rightarrow$ FEE Board Input $\rightarrow$ FEE Board Output $\rightarrow$ 10~dB Attenuator $\rightarrow$ VNA Port~2.

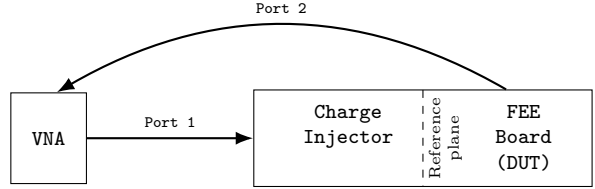
\begin{figure}[htbp!]
	\centering
        \begin{tikzpicture}[
            node distance=2.2cm,
            block/.style={
                rectangle, draw, minimum width=1cm, minimum height=1.2cm, align=center
            },
            arrow/.style={-{Latex}, thick},
            font=\ttfamily\scriptsize
            ]
            
            \node[block] (vna) {VNA}; 
            
            \node[block, right=of vna,
                text width=4.2cm,
                align=center,
                path picture={
                    \draw[dashed]
                        ([xshift=0cm]path picture bounding box.north) -- 
                        ([xshift=0cm]path picture bounding box.south);
                    \node[rotate=90, anchor=center, font=\tiny]
                        at ([xshift=0.3cm]path picture bounding box.center) {Reference\\plane};
                }
            ] (readout) {
                \begin{tabular}{p{1.8cm} p{2cm}}
                \centering Charge\\Injector & \centering FEE\\Board\\(DUT)
                \end{tabular}
            };
        
            \draw[arrow] (vna) -- node[above]{\tiny Port 1} (readout);
        
            \draw[arrow] 
                ([xshift=1.1cm]readout.north) 
                to[out=150, in=30] 
                node[above]{\tiny Port 2} 
                ([xshift=-0.4cm]vna.north east);
        
        \end{tikzpicture}
	\caption{Experimental setup for the frequency response analysis. The SOLT calibration was performed at the Reference Plane using a custom adapter. For clarity, the protective attenuators used during the measurement are not shown in this diagram. }
	\label{fig:exp-setupB}
\end{figure}

A two-port Short-Open-Load-Through (SOLT) calibration \cite{ShortOpenLoadThroughSOLTCalibration} was performed to de-embed the measurement chain, including the attenuators. For the \textit{Through} standard, a custom adapter was fabricated to connect the charge injector's input directly to an SMA output. Therefore, the VNA was calibrated to set its reference plane at the DUT's input and output, with the internal calibration algorithm accounting for the losses of the attenuators. No manual mathematical corrections were applied to the data post-measurement  (see Fig.~\ref{fig:exp-setupB}).

\subsection{Results}
\begin{figure}[H]
    \centering
    \includegraphics[width=10cm]{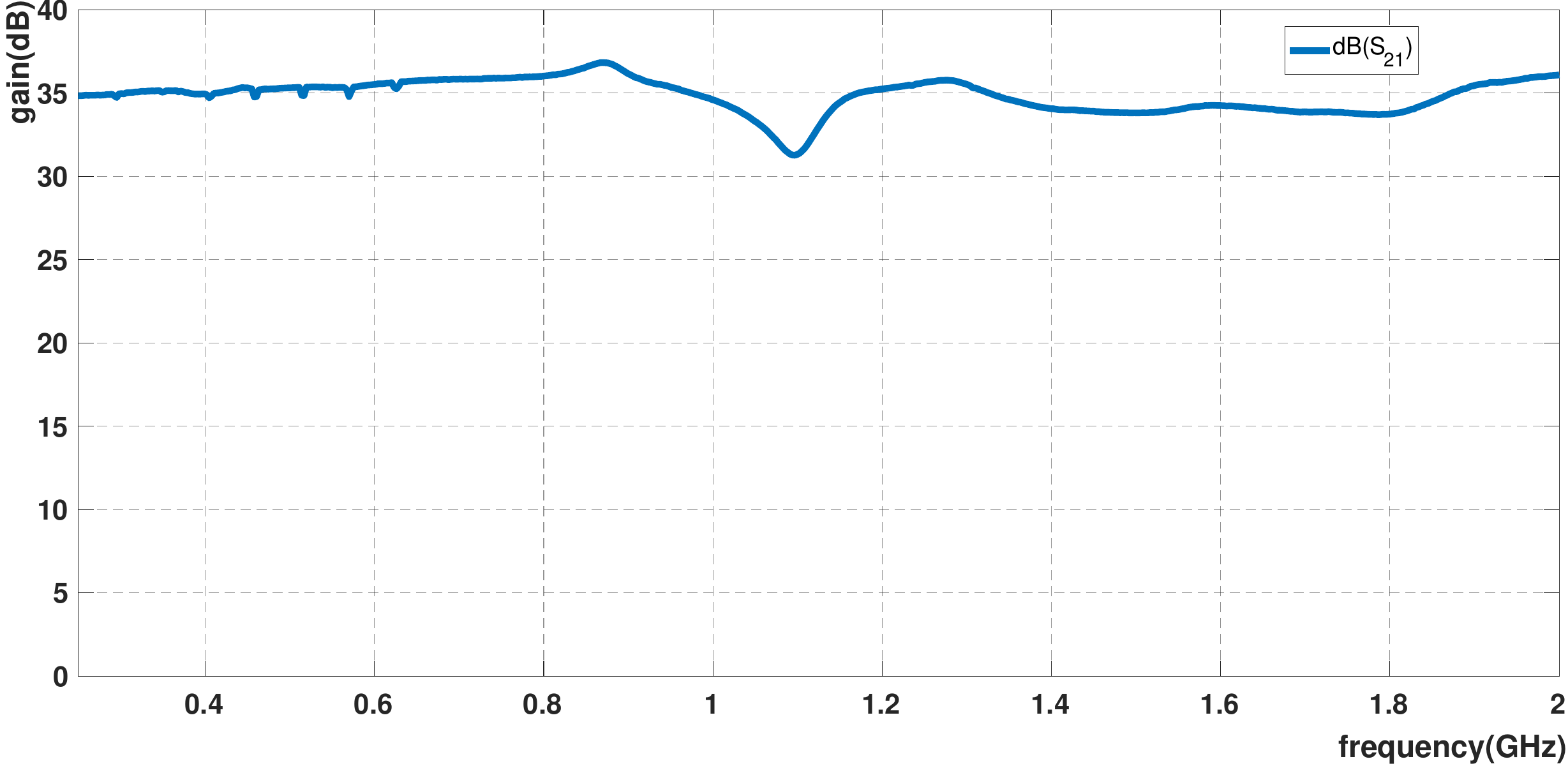}
    \caption{$S_{21}$ parameter for channel $8$ obtained with VNA. The gain is approximately $35\: dB$ for all the measured frequency range. Results are similar to table \ref{tab:gain_dB}.}
    \label{fig:s21_ch8}
\end{figure}

The forward voltage gain $S_{21}$ is plotted for channel $8$ in Fig.~\ref{fig:s21_ch8}. The gain is approximately $35\: dB$ across the measured frequency range. This value is systematically lower than the $\sim 38\: dB$ gain obtained with the pulse method (Table~\ref{tab:gain_dB}). This discrepancy is likely attributable to the systematic uncertainty introduced by the VNA calibration process itself. The use of a custom-fabricated, non-ideal \textit{Through} standard can lead to an imperfect de-embedding of the measurement chain by the VNA's algorithm, resulting in a systematic offset in the final $S_{21}$ measurement.

\section{Noise Spectrum}
\subsection{Experimental Setup}
We used the Spectrum Analyzer \textit{WR6ZI-RK-SPECTRUM} optional software package of the oscilloscope to estimate the noise spectrum introduced by the DUT. 

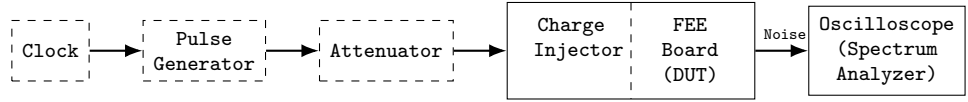
\begin{figure}[htbp!]
	\centering
        \begin{tikzpicture}[
            node distance=0.7cm,
            block/.style={rectangle, draw, minimum width=1cm, minimum height=0.8cm, align=center},
            dashedblock/.style={rectangle, dashed, draw, minimum width=1cm, minimum height=0.8cm, align=center},
            arrow/.style={-{Latex}, thick},
            font=\ttfamily\scriptsize
            ]
        
            \node[dashedblock] (clock) {Clock}; 
            \node[dashedblock, right=of clock] (pulse) {Pulse\\Generator};
            \node[dashedblock, right=of pulse] (attenuator) {Attenuator};
        
            \node[block, right=of attenuator,
                text width=3cm,
                align=center,
                path picture={
                    \draw[dashed]
                        ([xshift=0cm]path picture bounding box.north) -- 
                        ([xshift=0cm]path picture bounding box.south);
                }
            ] (readout) {
                \begin{tabular}{p{1cm} p{1.3cm}}
                \centering Charge\\Injector & \centering FEE\\Board\\(DUT)
                \end{tabular}
            };
        
            \node[block, right=of readout] (oscilloscope) {Oscilloscope \\ (Spectrum \\ Analyzer)};
        
            \draw[arrow] (clock) -- (pulse);
            \draw[arrow] (pulse) -- (attenuator);
            \draw[arrow] (attenuator) -- (readout);
            \draw[arrow] (readout) -- node[pos=1,above]{\tiny Noise \hspace*{0.5cm}} (oscilloscope);
        
        \end{tikzpicture}
	\caption{Experimental setup for the noise spectrum analysis. To estimate the baseline noise, all the system was connected but the FEE Board was not powered. To get the total output noise, all the setup was connected but no pulse signal was generated. Thus, the FEE Board noise was obtained according to equation \ref{eq:noiseest}.}
	\label{fig:exp-setupC}
\end{figure}

As explained in \cite{vaseghiAdvancedDigitalSignal2001}, the estimated magnitude spectrum $|\hat{X}(f)|$ could be extracted by subtracting the time-averaged noise $|\overline{N(f)}|$ from the signal magnitude spectrum $|Y(f)|$:
\begin{equation}
    |\hat{X}(f)| = |Y(f)| - |\overline{N(f)}|
    \label{eq:noiseest}
\end{equation}
In our case, for equation \ref{eq:noiseest}, $|\hat{X}(f)|$ is the FEE Board introduced noise spectrum, $|Y(f)|$ is the total output noise, and $|\overline{N(f)}|$ is the baseline noise introduced by the oscilloscope itself and other setup elements. This can be shown since the oscilloscope baseline noise is uncorrelated with the output noise from the FEE Board, thus we can apply the expectation operator and obtain:
\begin{eqnarray}
    \mathbb{E}[\mid \hat{X}(f)\mid ] &=& \mathbb{E}[\mid Y(f)\mid ] - \mathbb{E}[\mid\overline{N(f)}\mid ] \nonumber \\
    &=& \mathbb{E}[\mid X(f)+N(f)\mid ] - \mathbb{E}[\mid \overline{N(f) }\mid] \nonumber \\
    &\approx& \mathbb{E}[\mid X(f)\mid]
    \label{eq:noise}
\end{eqnarray}    
Therefore, in order to obtain the output noise power spectrum introduced by the board, we subtracted the oscilloscope baseline noise spectrum (all the system setup was connected, but the FEE readout board was not powered, see Fig.~\ref{fig:exp-setupC}) from the noise spectrum obtained as output from the FEE board powered on, without any input signal going through the system (see Fig.\ref{fig:noise_spectrum}).

\subsection{Results}
\begin{figure}[H]
    \centering
    \includegraphics[width=12cm]{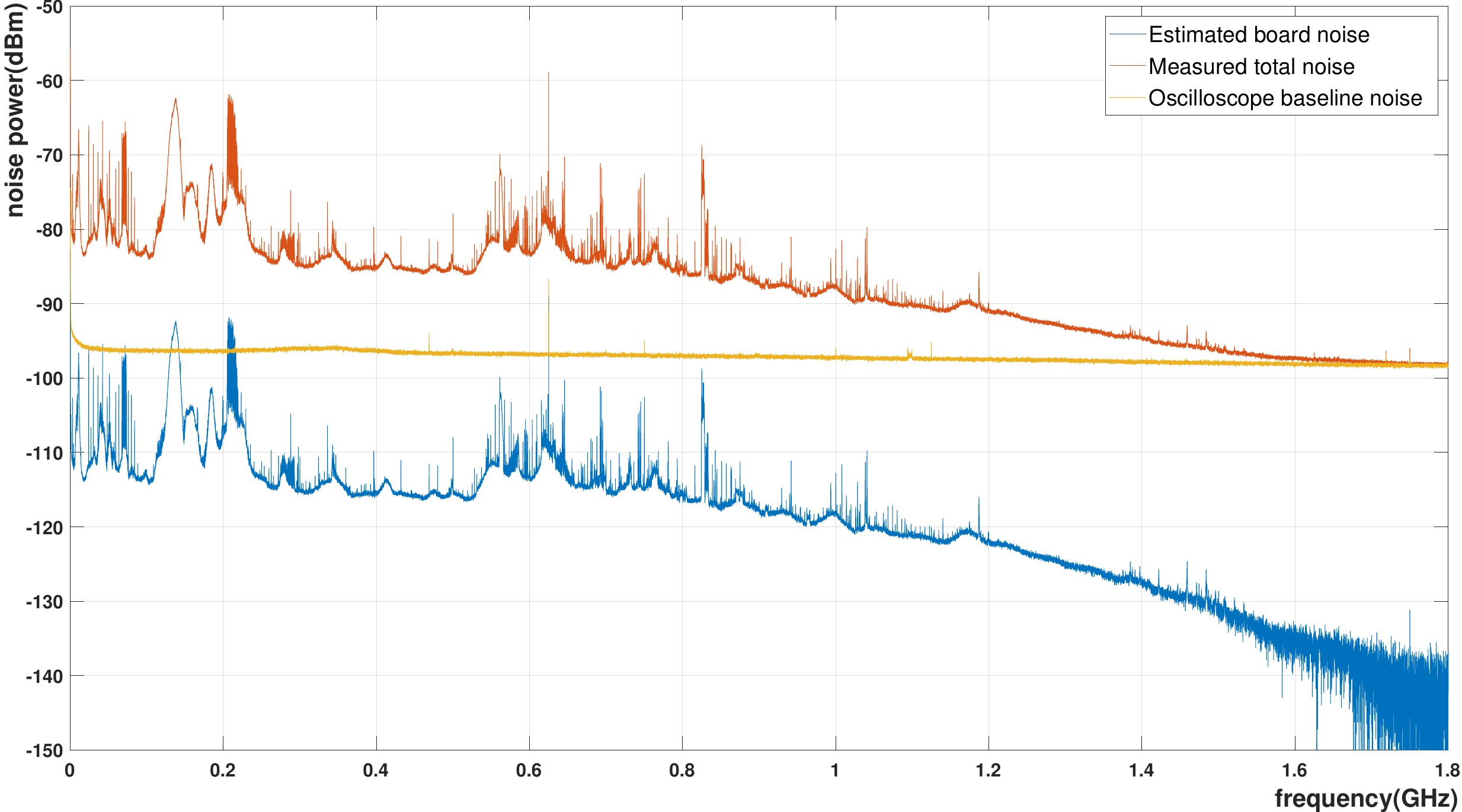}
    \caption{Output noise power spectra. The oscilloscope analog bandwidth is $1\: GHz$, which explains the gradual drop in the observed power spectra.}
    \label{fig:noise_spectrum}
\end{figure}

Figure~\ref{fig:noise_spectrum} shows the power spectral density of the noise at the output of the FEE board, after subtracting the baseline noise contribution from the oscilloscope and other setup elements. The estimated board noise (blue curve) is observed to be relatively flat across the frequency spectrum up to approximately 1~GHz. This is consistent with the nominal bandwidth of both the front-end electronics and the oscilloscope used for the measurement. The gradual drop in the observed power spectrum above 1~GHz is a direct consequence of the oscilloscope's analog bandwidth limit. Several discrete peaks are visible in the total measured noise, which are likely due to ambient electromagnetic interference in the laboratory environment not fully suppressed by the shielding enclosure. The integrated noise over this bandwidth \textit{should theoretically correspond} to the noise RMS values presented in Table~\ref{tab:noise-RMS}, which are a key parameter in determining the time resolution.

\section{Conclusions and Outlook}
We performed a detailed characterization of the front-end electronics (FEE) designed for the signal processing of AC-LGADs, using test signals similar in nature to those employed at FNAL for readout board evaluation. The time resolution of the experimental setup exhibited an asymptotic trend with increasing SNR, which allowed us to extract the contribution of the FEE to the overall time resolution.

The results confirm that the high-gain, low-noise amplification provided by the FEE is essential for achieving high time resolution. By preserving the initial signal-to-noise ratio of the sensor's signal while raising its amplitude far above the noise floor of the measurement system, the FEE enables precise timing measurements. Furthermore, the baseline noise is not significantly amplified; in fact, the observed noise levels fluctuate within a range that is approximately 10\% of the amplitude of the smallest signal generated in our setup.

The measured performance of the FEE meets the requirements established by previous simulation studies on LGAD pulse processing~\cite{penaSimulationModelFrontend2019}, which indicate that for $50\:\mu$m LGADs targeting a time resolution of $35$~ps, the FEE must provide a bandwidth exceeding $350$~MHz and ensure an SNR greater than 30.

This work focused on the characterization of a single channel. A complete assessment of the board's performance in a multi-channel environment would require the characterization of inter-channel effects, such as crosstalk. Such measurements necessitated the design and fabrication of a dedicated multi-channel charge injection board, which was beyond the scope of this initial study focused on establishing a robust methodology for time resolution analysis. This comprehensive inter-channel characterization remains a crucial step for future work.


\section{Acknowledgment}
We would like to thank Lautaro Narváez from the CALTECH INQNET team for the insightful discussions on the characterization.

This document was prepared using the resources of the Fermi National Accelerator Laboratory (Fermilab), a U.S. Department of Energy, Office of Science, HEP User Facility. 
Fermilab is managed by Fermi Research Alliance, LLC (FRA), acting under Contract No. DE-AC02-07CH11359.
This work was supported by the Chilean ANID - Millennium Science Foundation - ICN2019\_044, ANID PIA/APOYO AFB230003, ANID Fondecyt grant 1241803, ANID Fondecyt grant 1241685 and DIDULS grant PTE22538513.

\appendix






\end{document}